\documentclass[runningheads]{llncs}
\usepackage{graphicx}
\usepackage[utf8]{inputenc} 
\usepackage{hyperref}       
\usepackage{url}            
\usepackage{booktabs}       
\usepackage{amsfonts}       
\usepackage{nicefrac}       
\usepackage{microtype}      
\usepackage{xcolor}         
\usepackage{cite}
\usepackage{color,colortbl}
\usepackage{subcaption}
\usepackage{adjustbox}
\usepackage{multirow}
\usepackage{amsmath}
\usepackage{algorithm}      
\usepackage{algpseudocode}  

\usepackage[percent]{overpic}
\usepackage{enumitem}
\usepackage{amssymb}

\begin{document}
%
\title{Diffusion Model Driven Test-Time Image Adaptation for Robust Skin Lesion Classification}

%
%

\author{
Ming Hu \and
Siyuan Yan \and
Peng Xia \and
Feilong Tang \and
Wenxue Li \\
Peibo Duan \and
Lin Zhang \and
Zongyuan Ge
}
\institute{Monash University, Melbourne, Victoria, Australia 
}
\maketitle              
\begin{abstract}

Deep learning-based diagnostic systems have demonstrated potential in skin disease diagnosis. However, their performance can easily degrade on test domains due to distribution shifts caused by input-level corruptions, such as imaging equipment variability, brightness changes, and image blur. This will reduce the reliability of model deployment in real-world scenarios. Most existing solutions focus on adapting the source model through retraining on different target domains. Although effective, this retraining process is sensitive to the amount of data and the hyperparameter configuration for optimization. In this paper, we propose a test-time image adaptation method to enhance the accuracy of the model on test data by simultaneously updating and predicting test images. We modify the target test images by projecting them back to the source domain using a diffusion model. Specifically, we design a structure guidance module that adds refinement operations through low-pass filtering during reverse sampling, regularizing the diffusion to preserve structural information. Additionally, we introduce a self-ensembling scheme automatically adjusts the reliance on adapted and unadapted inputs, enhancing adaptation robustness by rejecting inappropriate generative modeling results. To facilitate this study, we constructed the ISIC2019-C and Dermnet-C corruption robustness evaluation benchmarks. Extensive experiments on the proposed benchmarks demonstrate that our method makes the classifier more robust across various corruptions, architectures, and data regimes. Our datasets and code will be available at \textit{\url{https://github.com/minghu0830/Skin-TTA_Diffusion}}.


\keywords{Skin lesion recognition  \and Test-time Domain Adaptation \and Diffusion Model.}
\end{abstract}
\section{Introduction}

\begin{figure}[t!]
\centering
\begin{overpic}[width=0.95\textwidth]{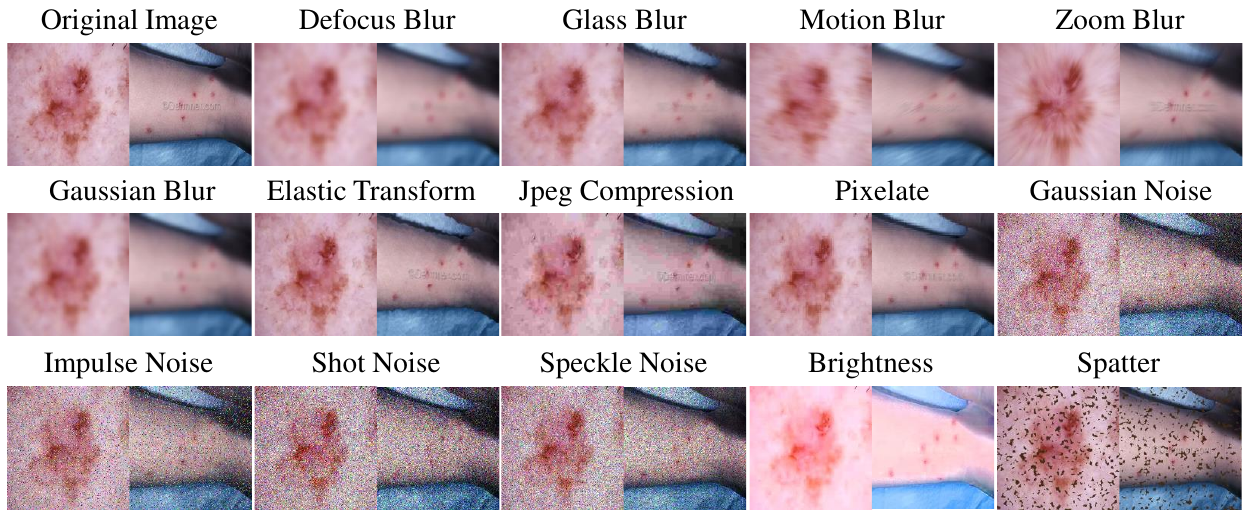}
\end{overpic}
\caption{The ISIC2019-C and Dermnet-C datasets we proposed contain 14 types of algorithmically generated corruptions from noise, blur, and digital categories, aiming to simulate the common appeared image-level corruption in real-world scenario. Each type of corruption has five levels of severity, resulting in 70 distinct corruptions. In each example, the images on the left are dermatoscopic images, and those on the right are clinical images captured by regular cameras.}
\label{fig:corruption_pre}
\vspace{-0.65 cm}
\end{figure}

The diagnosis and treatment of skin diseases heavily rely on visual cues, making dermatology a field well-suited for the application of computer-aided diagnosis systems (CAD), particularly deep learning techniques~\cite{chen2022efficient, xia2023hgclip, xia2023lmpt}. Despite significant advances, the performance of deep learning models in dermatology is often degraded due to distribution shifts on test data. Previous works \cite{yan2023epvt, yan2024prompt, bissoto2022artifact} have identified and addressed distribution shift issues of skin lesion classifiers caused by dermoscopic artifacts such as dark corners, rulers, and surgical markings. However, an important yet underexplored distribution shift is image-level corruption, such as changes in brightness, blurring, pixelation, skin color and combinations thereof, which can often lead to unreliable diagnoses~\cite{Guan_2022, hendrycks2019robustness, zheng2024benchmarking, Yan_2023_CVPR, ming2023nurvid}, as show in Fig.~\ref{dataset_sta2}. These corruptions are common in skin images due to imaging equipment variability and low-lighting conditions, as shown in Fig.~\ref{fig:corruption_pre}.

To address the image-level corruption issue in skin lesion classifiers, domain adaptation becomes essential to ensure optimal performance under distribution shifts~\cite{yu2023comprehensive}. Traditional domain adaptation~\cite{ganin2015unsupervised, long2016unsupervised,motiian2017unified} aims to align training and test data representations in the embedding space to achieve domain-invariant representations. In the current adaptation paradigm, models are refined during training through joint optimization across source and target domains. However, this approach faces a critical limitation: the unpredictability of data variation at the testing stage. Although updates during training can address shifts within known target domains, unforeseen variations during deployment present a significant challenge. To mitigate this, test-time model adaptation becomes essential, enabling model adjustments without source data and without interrupting ongoing inference processes.  Simultaneously, test-time model adaptation strategies refine the model during prediction phases, enhancing resilience to domain shifts. Despite these advancements, real-time model updates introduce complexities, particularly in computational demands, which may not be feasible for scaling across multiple targets. Each target might necessitate a unique model adaptation, sensitive to varying data quantities or sequences, potentially undermining model robustness.

In this paper, we propose a novel test-time image adaptation method that overcomes the limitations of existing test-time model adaptation methods, such as the need for fine-tuning the classifier model during testing and the requirement for paired source and target data for training.
Our proposed diffusion-based method innovates by updating target data during testing through generative diffusion modeling. 
During the training process, we train a diffusion model on source domain data and a discriminative classification model on source domain data and annotations. In the testing phase, given a test input from the target domain, the diffusion model projects it to the source domain.We adopt an iterative latent refinement process, which conditions on the input image in the reverse process. To enhance structural and class alignment between the generated and input images, we design a structured guidance module based on linear low-pass filters, performing a series of downsampling and upsampling operations by a scale factor. In each reverse time step $t$, potential refinement operations are added, ensuring that the low-pass filtered image represents the structural information of the image. Additionally, we employ a confidence-based self-ensembling scheme to aggregate predictions for the original and adapted inputs, automatically selecting the degree of reliance on adapted and unadapted inputs. This approach enhances adaptation robustness by rejecting inappropriate results from generative modeling. These enable both source-free and test-time adaptation. Our contributions can be summarized as:

\begin{figure*}[t!]
\centering
\subfloat[original test set]{
    \includegraphics[width=0.32\textwidth, trim=50 50 50 50, clip]{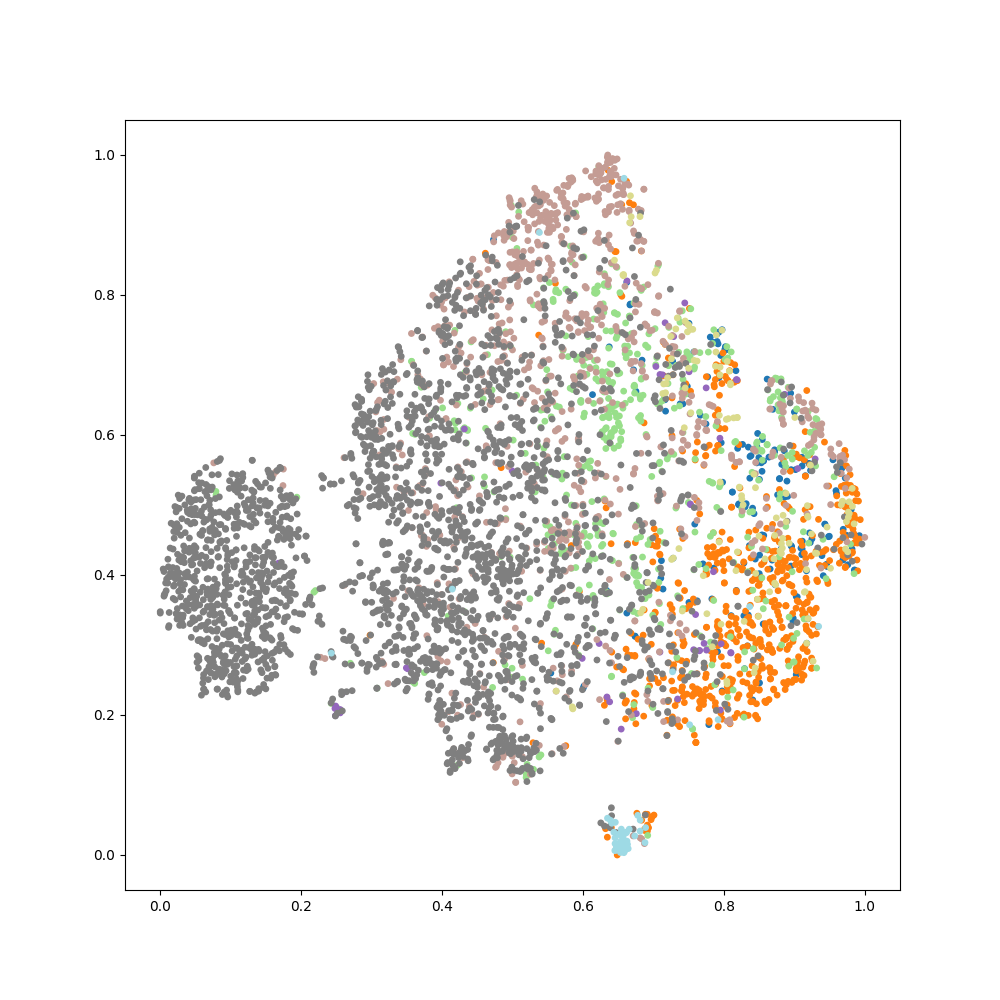}
}
\subfloat[corrupted test set]{
    \includegraphics[width=0.32\textwidth, trim=50 50 50 50, clip]{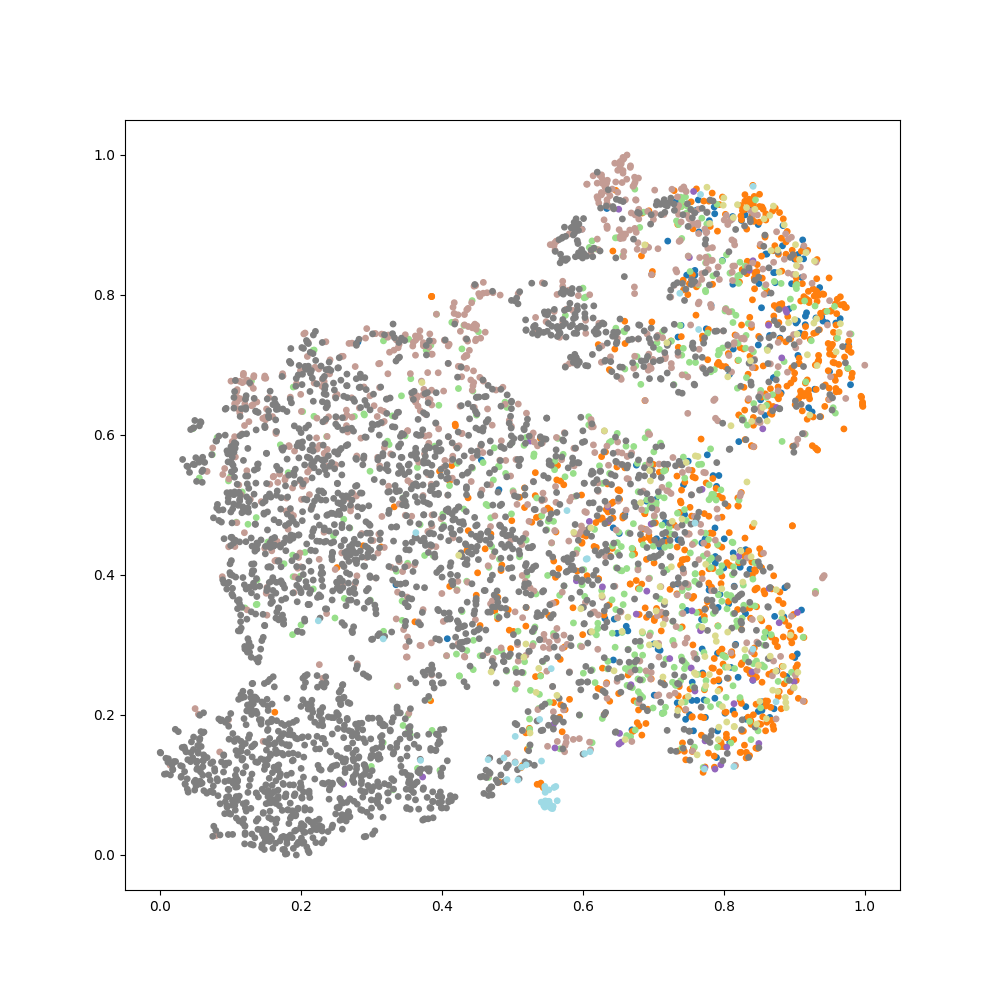}
}
\subfloat[generated test set]{
    \includegraphics[width=0.32\textwidth, trim=50 50 50 50, clip]{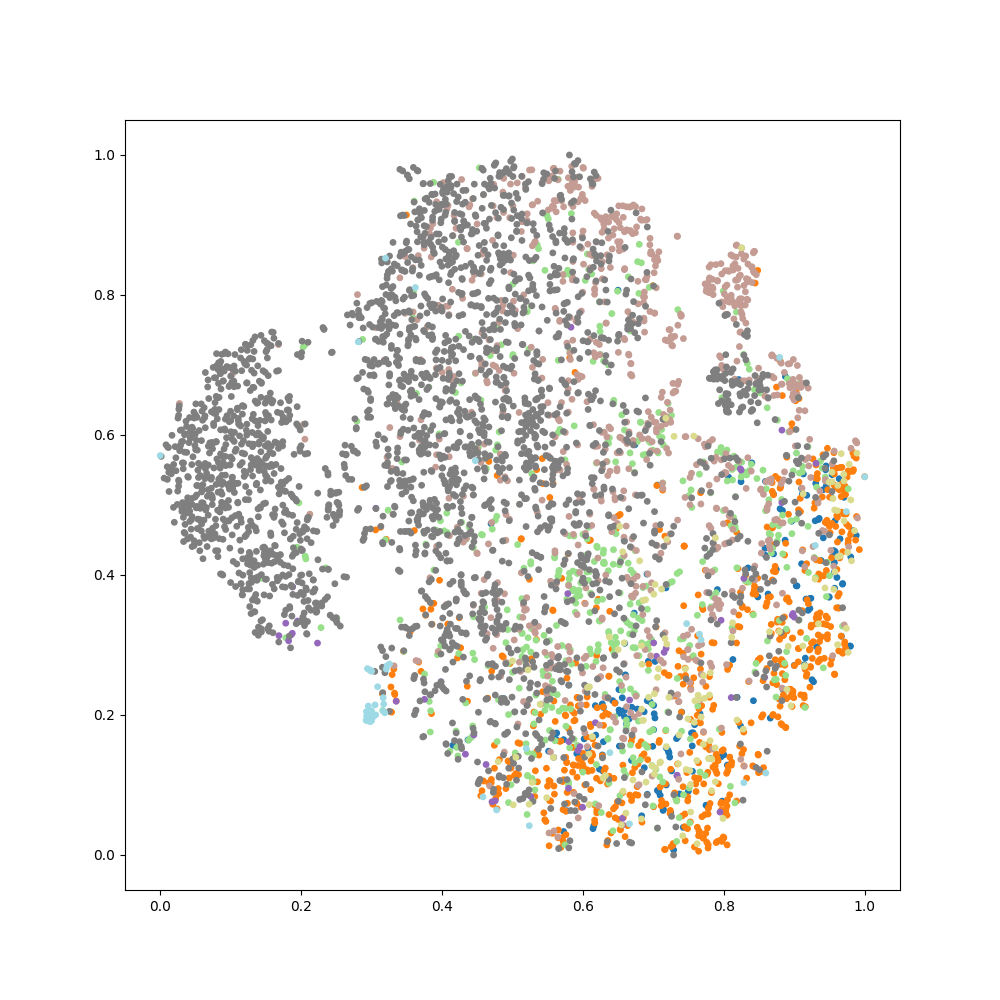}
}
\caption{We show t-SNE plots of the original ISIC 2019 test set, a corrupted test set with gaussian noise (ISIC2019-C dataset, severity = 3), and test set restored to the target domain using our diffusion-driveen method. The classifier is Swin-T. As shown in the figure, the classifier experiences a significant drop in robustness on corrupted images, along with the improvement effect of our method.}
\label{dataset_sta2}
\vspace{-0.50cm}
\end{figure*}

\begin{itemize}[leftmargin=*]
\item[$\bullet$] 
We constructed the ISIC2019-C and Dermnet-C corruption robustness evaluation benchmarks, aims to simulating the common appeared image-level corruption in skin images in real world scenario. We introduce a total of 14 corruption types including noise, blur, and digital forms. Each type of corruption has five levels of severity, resulting in 70 distinct corruptions. To our best knowledge, this is the first work to comprehensively evaluate the corruption robustness of dermatological images. 


\item[$\bullet$]
We propose a training-free and diffusion-driven test-time image adaptation method, featuring a novel structured guidance module that enhances structural and class alignment. A confidence-based self-ensembling scheme balances corrupted and adapted inputs. Experiments demonstrate that our approach improves robustness without altering the model itself or supplementing target domain data.

\end{itemize}

\section{Methodology}
In our approach, we train a class-unconditional diffusion model on source domain data for test-time image adaptation and a discriminative classification model on annotated source data. We employ an iterative latent refinement process with a structured guidance module based on linear low-pass filters to enhance structural and class alignment in the reverse diffusion process. To address diffusion failure and enhance robustness, we utilize a confidence-based self-ensembling scheme that automatically balances reliance on original and adapted inputs, improving adaptation robustness by rejecting unsuitable generative modeling outcomes.

\subsection{Background: Class-unconditional Diffusion Modeling}

Diffusion models~\cite{sohl2015deep, rombach2022high, song2020improved, song2020score} have recently achieved state-of-the-art image generation by iteratively refining noise into samples from the data distribution. A diffusion model consists of three major components: the forward process, the reverse process, and the loss function. 

\noindent \textbf{Forward Process.} Given an image sampled from the real data distribution $x_0 \sim q(x_0)$, the forward diffusion process defines a fixed Markov chain, to gradually add Gaussian noise to the image $x_0$ over $T$ timesteps,
producing a sequence of noised images $x_1, x_2, \cdots, x_T$. The forward process can be defined as:

\begin{equation}\label{eq:forward_x0}
	\begin{split}
		&q(x_{1:T}|x_{0}) := \prod_{t=1}^{T}q(x_{t}|x_{t-1}), \\
		& q\left(x_{t} \mid x_{t-1}\right):=N\left(x_{t} ; \sqrt{1-\beta_{t}} x_{t-1}, \beta_{t} \mathbf{I}\right),
	\end{split}
\end{equation}
where the sequence, $\beta_{1}, ..., \beta_{T}$, is a fixed variance schedule to control the step sizes of the noise.

We can obtain by deparameterization:
\begin{equation}\label{eq:forward}
    q(x_t|x_0) := \sqrt{\bar{\alpha}_t}x_0+\epsilon\sqrt{1-\bar{\alpha}_t}, \epsilon \sim \mathcal{N}(0, 1),
\end{equation}
where $\alpha_{t} := 1 - \beta_{t}$ and $\overline{\alpha_{t}} := \prod_{s=1}^{t}\alpha_{s}$.

\noindent \textbf{Reverse Process.} The reverse diffusion process iteratively removes the noise to generate an image in $T$ timesteps:
\begin{equation}\label{eq:reverse1}
	\begin{split}
		&p_{\theta}\left(x_{t-1} \mid x_{t}\right):=N\left(x_{t-1} ; \mu_{\theta}\left(x_{t}, t\right), \sigma_{t}^{2}\left(x_{t}, t\right) \mathbf{I}\right).
	\end{split}
\end{equation}

Where the mean and variance are both calculated according to the model. Combining all time steps, we can obtain:
\begin{equation}\label{eq:reverse2}
	\begin{split}
		&p(x_{0:T}) := p(x_T)\prod_{t=1}^{T}p(x_{t-1}|x_{t})
	\end{split}
\end{equation}

\begin{figure}[t!]
\centering
\begin{overpic}[width=0.95\textwidth]{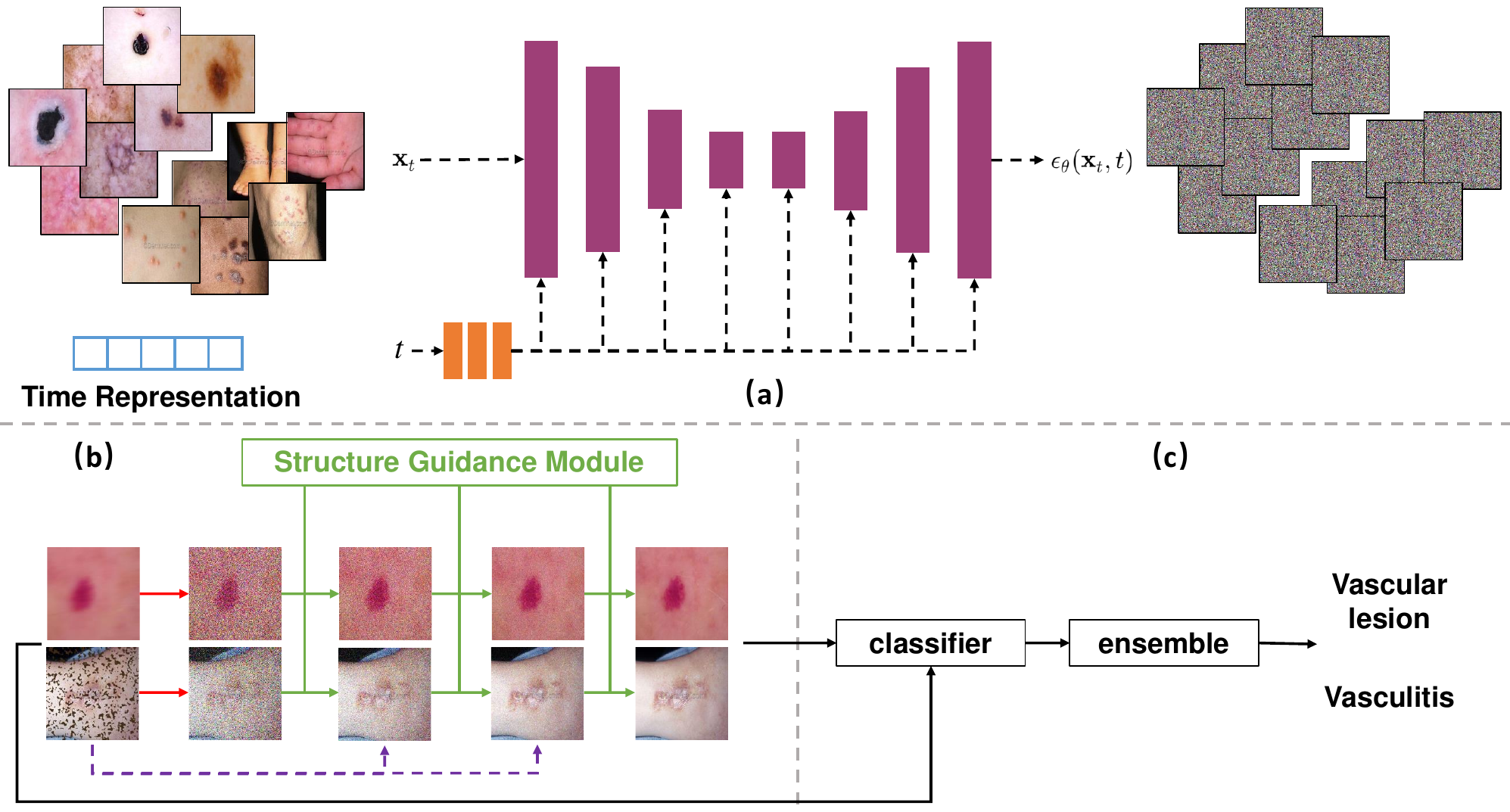}
\end{overpic}
\caption{Our approach is based on a diffusion model (Subfig.(a)) that projects the target input back to the target domain. Adapting the input during testing enables direct use of the source classifier without the need for model adaptation. The projection adds noise (Subfig.(b), forward diffusion, red arrow), then iteratively updates the input with conditioning on the original input (reverse diffusion, purple arrow) (Subfig.(b), guidance, green arrow). For reliability, we ensemble predictions (Subfig.(c)) with and without adaptation depending on their confidence.}
\label{fig:model_arch}
\vspace{-0.55 cm}
\end{figure}

\noindent \textbf{Loss Function.} 
Denoising diffusion probabilistic models (DDPM)~\cite{ho2020denoising} introduce a time-dependent constant $\sigma_{t}\left(x_{t}, t\right)=\sigma_t\mathbf{I}$ for diffusion. The model parameterizes $\mu_{\theta}\left(x_{t}, t\right)$ as a linear combination of $x_{t}$ and a noise prediction function $\epsilon_{\theta}(x_{t}, t)$. The optimization of $\mu_{\theta}\left(x_{t}, t\right)$'s parameters aims to minimize the variational bound on the negative log-likelihood $\mathbb{E}[-\log p_\theta(x_0)]$. Following DDPM, the training loss $\mathcal{L}_{\text{simple}}$ is reduced to the mean-squared error between the actual noise $\epsilon \sim \mathcal{N}(0, \mathbf{I})$ in $x_t$ and its estimation:

\begin{equation}\label{eq:loss}
\mathcal{L}_{\text{simple}} := ||\epsilon_{\theta}(x_{t}, t) - \epsilon||^2.
\end{equation}

Diffusion models are optimized to learn a generative prior of the training data, as their loss is derived from a bound on the negative log-likelihood $\mathbb{E}[-\log p_\theta(x_0)]$.

\subsection{Structure Guidance Module}

Diffusion models face a trade-off between preserving class information and aligning domains when projecting target images to the source domain. Too much noise obscures categories, while insufficient noise fails to transform domains. Lesion boundary shapes also vary across samples, crucial for diagnosis, increasing transformation difficulty. Inspired by ILVR~\cite{choi2021ilvr}, we introduce a structured guidance module based on linear low-pass filters, $\phi_D(\cdot)$, performing downsampling and upsampling to capture lesion structures. We iteratively update the diffused sample $x^g_t$ to reduce structural differences measured by $D$ between generated and input images. At each reverse step, we obtain estimates of the original $x_0$ and transformed $\hat{x}^g_0$ from the current noisy $x^g_t$:



\begin{equation}\label{eq:x0}
\hat{x}^g_0 = \sqrt{\frac{1}{\bar{\alpha}_t}} \left( x^g_t - \sqrt{1 - \bar{\alpha}t} \boldsymbol{\epsilon}\theta(x_t^g, t) \right).
\end{equation}

This approach balances preserving category information and aligning domains while retaining critical lesion boundaries. To avoid conflicting with the diffusion update, we use the direction of similarity between the reference image $x_0$ and the transformed estimate $\hat{x}^g_0$, instead of the one between $x_t$ and $x^g_t$. At each step $t$ in the reverse process, we force $x^g_t$ to move in the direction that decreases the distance between $\phi_D(x_0)$ and $\phi_D(\hat{x}^g_0)$, with a scaling hyperparameter $\boldsymbol{w}$ to control the step size of guidance:

\begin{equation}\label{eq:guidance}
    x^g_{t-1} = \hat{x}_{t-1}^{g} - \boldsymbol{w} \nabla_{x_t}\left\|\phi_{D}\left(x_{0}\right) -\phi_{D}\left(\hat{x}^g_0\right)\right\|_2,
\end{equation}
with a scaling hyperparameter $w$ to control the step size of guidance.
\subsection{Confidence-based Self-Ensembling}
By adjusting the target inputs back to the source domain through diffusion model, our method enable our source-trained classifier to make more reliable predictions on the target domain. The images generated by diffusion often improve accuracy as they retain class information while projecting the target input towards the source domain. However, the diffusion model can sometimes generate images more challenging for the classifier than the original targets.  To address this, we employ a confidence-based self-ensembling scheme. We select the classification result with the highest confidence, i.e., $\arg\max_c\left(\mathrm{if~}\max(p)>\max(p^g)\mathrm{~then~}p_c\mathrm{~else~}p_c^g\right)$, $\operatorname{where}c\in\{1,\ldots,C\}$, and the confidence of the $C$ categories is $p\in\mathbb{R}^C$ and $p^g\in\mathbb{R}^C.$ This approach automatically chooses the more reliable prediction between original and diffused inputs, enhancing robustness by rejecting inappropriate results from generative diffusion modeling when necessary.




\begin{figure}[t!]
\centering
\begin{overpic}[width=0.95\textwidth]{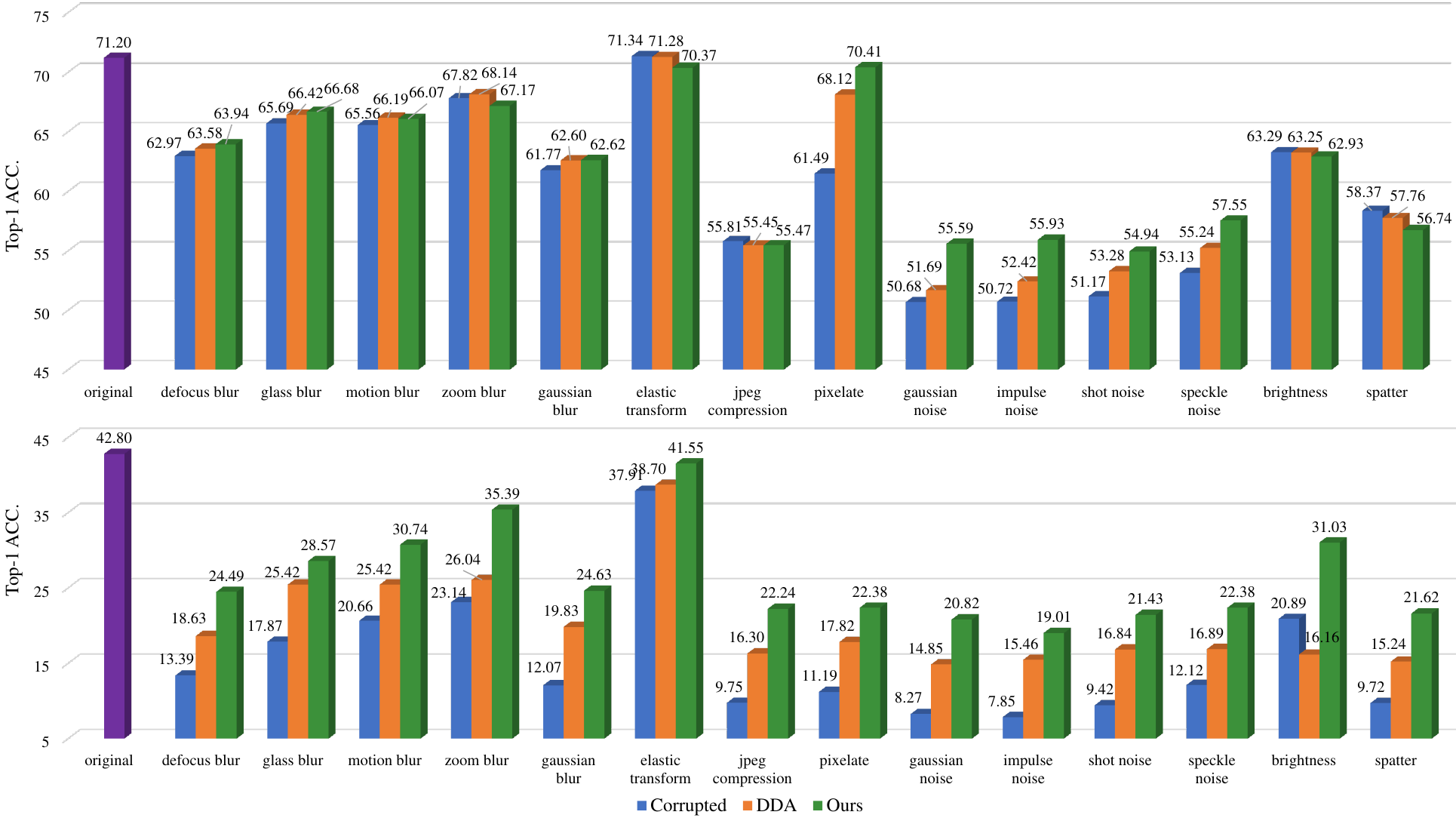}
\end{overpic}
\caption{We demonstrate the performance of Swin-T across all corruptions (severity = 5). Our method reliably improves classifier robustness across most corruption types, especially on the Dermnet-C dataset (bottom panel). Images captured by general cameras are more perturbed compared to those captured by dermatoscopes, and our method prevents catastrophic drops.}
\label{fig:data_compare}
\vspace{-0.65 cm}
\end{figure}

\section{Experiments}
\noindent \textbf{Datasets.} The corruptions are drawn from three main categories—noise (gaussian noise, impulse noise, shot noise, and speckle noise), blur(defocus blur, glass blur, motion blur, zoom blur, and gaussian blur), digital (elastic transform, jpeg compression, and pixelate) and extra (brightness and spatter), as shown in Fig.~\ref{fig:corruption_pre}. For the construction of ISIC2019-C, we redivided the ISCI 2019 dataset into training, validation, and test sets in a ratio of 0.7:0.1:0.2, comprising a total of 25,331 dermatoscopic images across eight different diagnostic categories: melanoma, melanocytic nevus, basal cell carcinoma, actinic keratosis, benign keratosis (solar lentigo, seborrheic keratosis, and lichen planus-like keratosis), dermatofibroma, vascular lesion, and squamous cell carcinoma. Dermnet-C, on the other hand, utilized the original Dermnet dataset split, consisting of 19,500 images, encompassing 23 skin conditions such as acne, melanoma, eczema, seborrheic keratosis, ringworm, pemphigus, poison ivy, psoriasis, and vascular tumors. It is important to note that Dermnet's images are not dermatoscopic but captured using mobile or handheld devices. Finally, we corrupted the test sets of these two datasets.
\\

\noindent \textbf{Training and Evaluation.}
To ensure a comprehensive evaluation, we experimented with multiple classification architectures. Specifically, we standardized on ResNet-50~\cite{he2016deep} as a common architecture and also evaluated the state-of-the-art in attentional and convolutional architectures using Swin-T~\cite{liu2021swin} and ConvNeXt~\cite{liu2022convnet}. 
We report the classifier performance on the original test set, as well as the robustness on corrupted test set. In terms of methodology, we compare with DDA~\cite{gao2022back}, a state-of-the-art method for test-time input domain adaptation model driven by the diffusion model on natural images. The test results are all reported as Top-1 accuracy. \\

\begin{table}[t!]
    \centering
    \caption{We compared the Top-1 accuracy of classifiers on the original test set, the corrupted test set, and images generated using our method.}
    \label{t1}
\begin{tabular}{l *{8}{>{\centering\arraybackslash}p{1.2cm}}}
    \toprule
        & \multicolumn{4}{c}{ISIC2019-C} & \multicolumn{4}{c}{Dermnet-C} \\
    \cmidrule(lr){2-5} \cmidrule(lr){6-9}
        Model & {Original} & {Corrupted} & {DDA} & {\textbf{Ours}} & {Original} & {Corrupted} & {DDA} & {\textbf{Ours}}\\
    \midrule
        \multirow{1}{*}{ResNet-50} & 66.54  &  55.33  &  56.18  & \textbf{56.72}  & 39.88 & 19.61 & 28.66 & \bf 31.56 \\
        \multirow{1}{*}{Swin-T} & 71.20  &  64.08  &  65.25  &  \textbf{66.04}  & 42.80 & 22.75 & 31.52 & \bf 37.55 \\
        \multirow{1}{*}{ConvNeXt-T} & 73.29 & 66.43  &  67.44  & \textbf{68.11} & 46.40  &  38.72 & 41.52 & \bf 43.32 \\
    \bottomrule
    \end{tabular}
\vspace{-0.65cm}
\label{table1}
\end{table}

\noindent \textbf{Results.}
In Tab.~\ref{table1}, we present the overall results. All classifiers exhibit a notable decline in accuracy on corrupted datasets. Both the DDA and our proposed methods enhance accuracy on these datasets, with our method consistently outperforming DDA across all models and datasets. 
For instance, our method improves the accuracy of ResNet-50, Swin-T, and ConvNeXt-T on corrupted sets by 1.39\%, 1.96\%, and 1.68\% respectively, compared to using DDA alone.
Fig.~\ref{fig:data_compare} shows the performance of the Swin-T classifier across all corruptions at severity level 5, illustrating that most corruptions negatively impact the classifier, with the effect proportional to the severity. Additional detailed experimental results are provided in the supplementary material. For example, Demrnet's performance drops from 42.80\% to 13.39\% after defocus blur corruption, but our proposed domain adaptation method improves it to 24.49\% without altering the classifier or using external domain data. \\

\noindent \textbf{Ablation Study.}
We conduct two ablative experiments: (1) exploring diffusion updates: a) forward+reverse without guidance; b) reverse+guidance starting from noise; and c) our full method combining both with guidance. (2) investigating ensemble methods: We assess hard selection and soft fusion based on prediction entropy/confidence. Hard selection chooses images with lower entropy (higher confidence), while soft fusion combines images or logits based on their entropy/confidence. All experiments are performed on the ISIC2019-C dataset.

\begin{table}[t!]
    \begin{minipage}[t]{0.47\linewidth}
        \centering
        \caption{Ablation of diffusion updates justifies each step.}
        \label{tab:diffusion_module_ablation}
        \adjustbox{max width=\linewidth}{
            \renewcommand\arraystretch{1.15}
            \begin{tabular}{ll|ccc}
                & & ResNet-50 & Swin-T & ConvNeXt-T \\
                \midrule
                \multirow{3}{*}{\rotatebox{90}{(a) noise}}
                & forward+reverse & 20.65 & 31.77 & 34.45 \\
                & reverse+guidance & 20.52 & 30.15 & 32.96 \\
                & Ours & \bf 40.89 & \bf 63.43 & \bf 63.33 \\
                \midrule
                \multirow{3}{*}{\rotatebox{90}{(b) blur}}
                & forward+reverse & 48.37 & 52.23 & 55.28 \\
                & reverse+guidance & 39.56 & 50.19 & 53.72 \\
                & Ours & \bf 65.38 & \bf 68.25 & \bf 71.35 \\
                \midrule
                \multirow{3}{*}{\rotatebox{90}{(c) digital}}
                & forward+reverse &  57.64 & 59.02  & 61.55  \\
                & reverse+guidance & 53.14  & 56.35  & 57.49  \\
                & Ours & \bf 65.20  & \bf 67.76  & \bf 70.26  \\
                \midrule
                \multirow{3}{*}{\rotatebox{90}{(d) else}}
                & forward+reverse &  52.86 & 63.67  & 64.55  \\
                & reverse+guidance &  51.77 &  60.30 &  62.54 \\
                & Ours & \bf 56.54  & \bf 66.05  & \bf 68.11  \\
                \bottomrule
            \end{tabular}
        }
    \end{minipage}
    \hfill
    \begin{minipage}[t]{0.48\linewidth}
        \centering
        \caption{Ablation on the design choices of selection module.}
        \label{tab:ablationImageNetC2in1}
        \adjustbox{max width=\linewidth}{
            \renewcommand\arraystretch{1.2}
            \begin{tabular}{l|ccc}
                & ResNet-50 & Swin-T & ConvNeXt-T \\
                \midrule
                original test & 66.54 & 71.20 & 73.29\\
                \midrule
                corruption & 55.33 & 64.08 & 66.43 \\
                diffusion & 56.54  & 64.88 & 67.25\\
                \midrule
                entropy fuse & 55.48 & 65.25 & 66.52\\
                confidence fuse & 55.47 & 65.20  & 66.60 \\
                \midrule
                entropy sum & 55.63 & 65.40 & 66.90 \\
                confidence sum & 55.63 &  65.25 & 67.50 \\
                sum & 56.18 & 65.25 & 67.44 \\
                 \midrule
                entropy & 56.20 & 64.98 & 67.24\\
                \rowcolor{lightgray}
                confidence (Ours) & \bf 56.72 & \bf 66.04  & \bf 68.11 \\
                \bottomrule
            \end{tabular}
        }
    \end{minipage}
\vspace{-0.45cm}
\end{table}

\section{Conclusion}
In conclusion, we propose a test-time image adaptation method to enhance the robustness of deep learning-based skin disease diagnosis systems under distribution shifts caused by input-level corruptions. By diffusing corrupted target images back to the source domain using a generative diffusion model, We designed a novel structured guidance module to enhance structural and class alignment, and used a confidence-based self-ensembling scheme to balance corrupted and adapted inputs. Our method improves the classifier's accuracy across various corruptions, architectures, and data regimes, as demonstrated on our proposed ISIC2019-C and Dermnet-C benchmarks. 

\noindent \textbf{Acknowledgment:} This work primarily explores the performance of diffusion models under the TTA setting for skin diseases, referencing the DDA framework.

\bibliographystyle{splncs04}
\bibliography{main}

\clearpage
\section{Appendix for ``Diffusion Model Driven Test-Time Image Adaptation for Robust Skin Lesion Classification"}

\begin{figure}[ht!]
    \centering
    \captionsetup{skip=5pt}
    \includegraphics[width=0.65\textwidth]{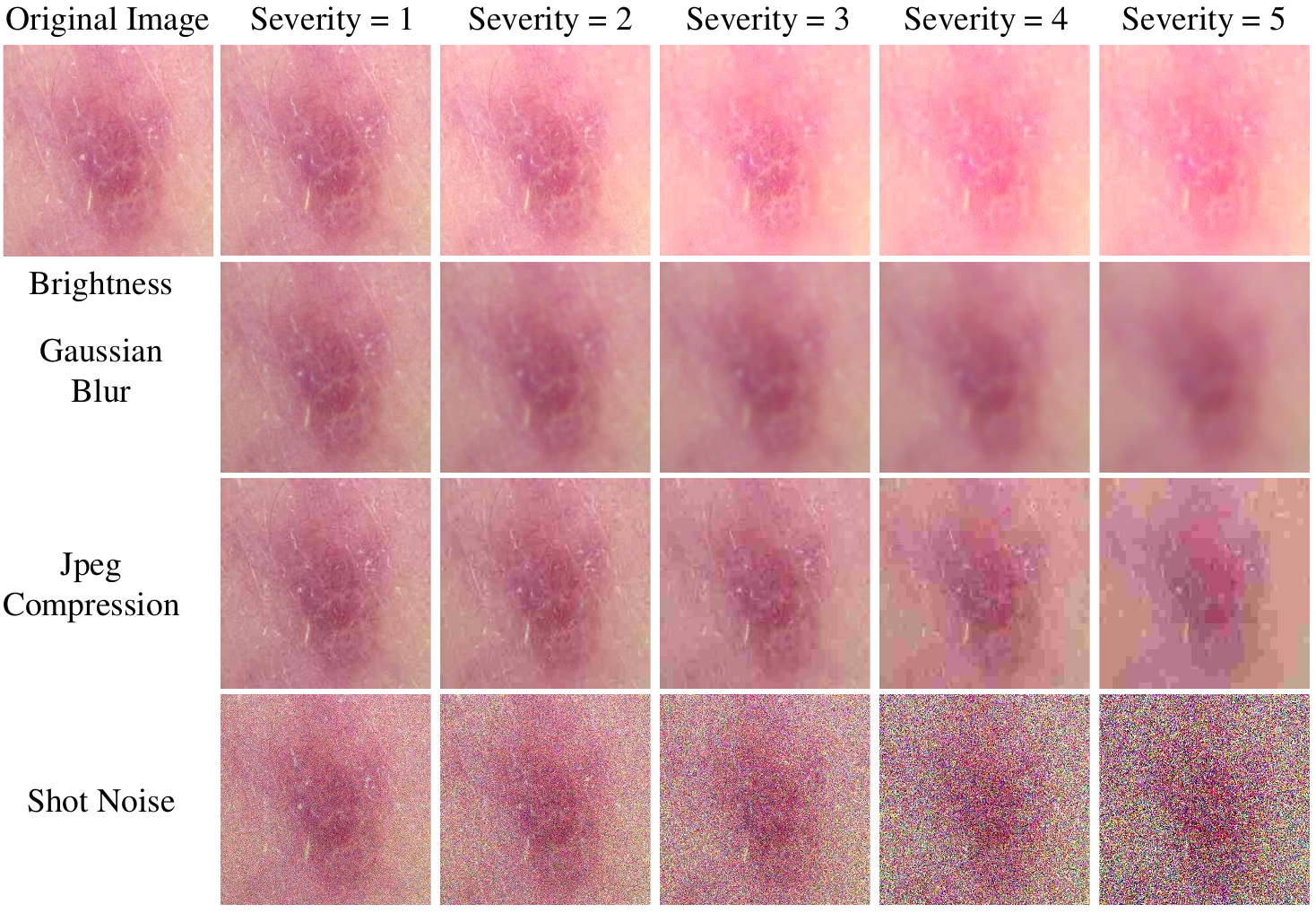}
    \caption{Examples of different corruptions at five different severity levels.}
    \vspace{0.1cm} 
    \includegraphics[width=0.65\linewidth]{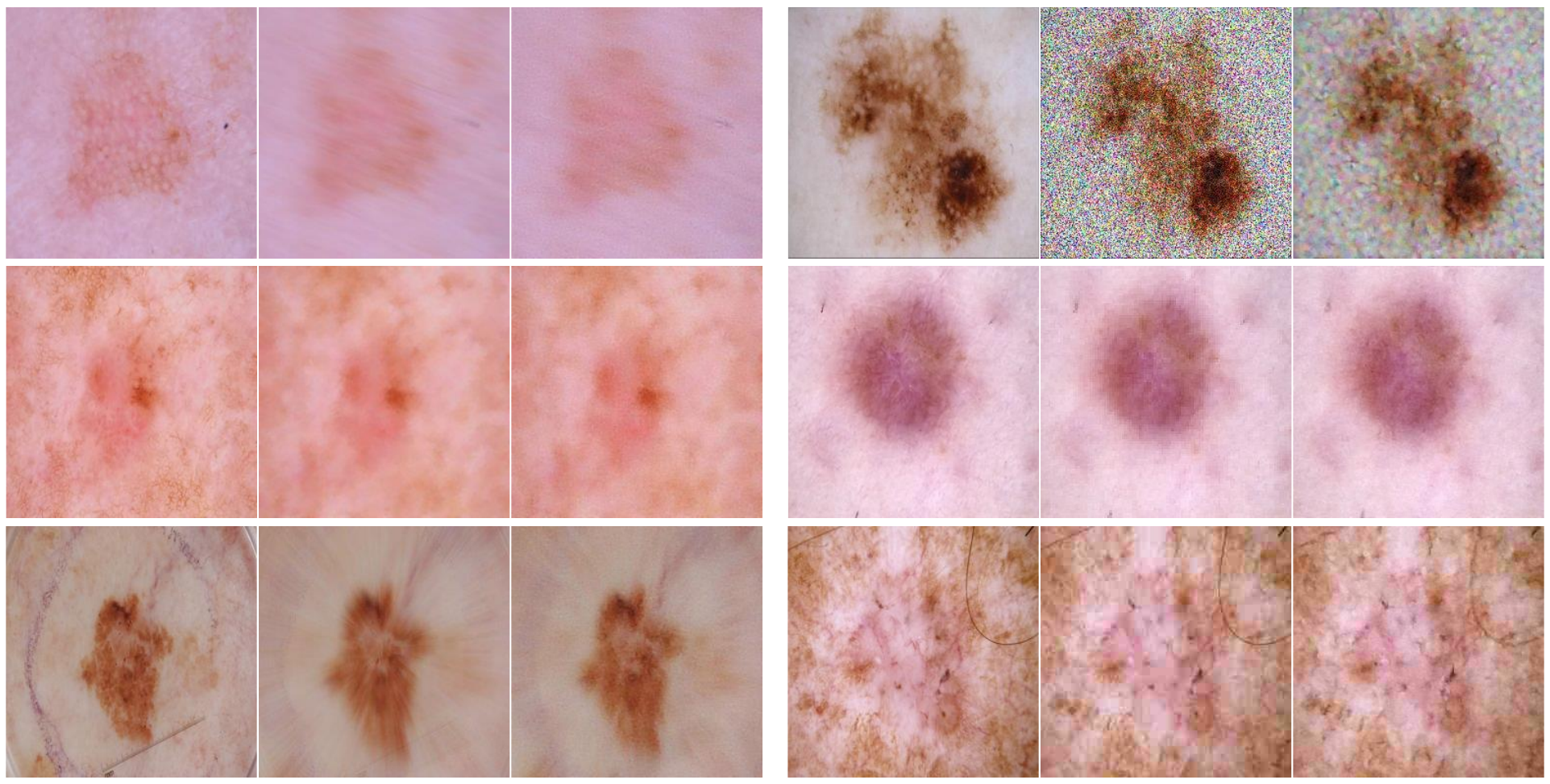}
    \vspace{0.1cm} 
    \includegraphics[width=0.677\linewidth]{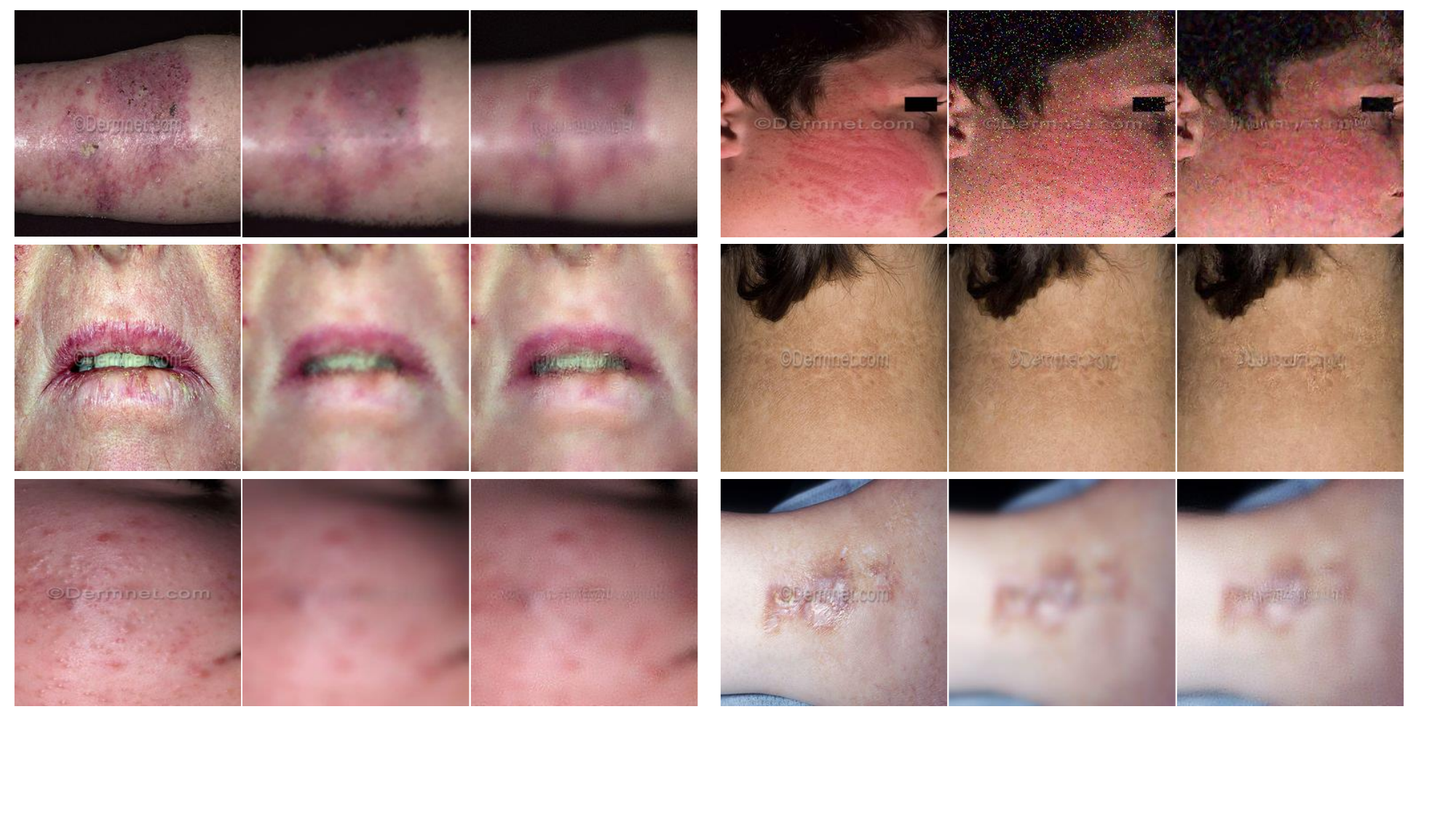}
    \hspace*{-0.425cm}
    \vspace{-0.5cm} 
    \caption{Some generated examples, \textbf{left:} original image, \textbf{middle:} corrupted image, \textbf{right:} generated image.}
\end{figure}

\begin{table}[t!]
  \tiny
  \centering
  \caption{Results of all corruptions and classifiers on the ISIC2019-C dataset.}
    \begin{tabular}{l|c|ccc|ccc|ccc|}
          & \multicolumn{1}{c}{} & \multicolumn{3}{p{12.57em}}{ResNet-50} & \multicolumn{3}{p{12.57em}}{Swin-T} & \multicolumn{3}{p{12.57em}}{ConvNeXt-T} \\
\cmidrule{2-11}          & \multicolumn{1}{p{5.5em}|}{Corruption level} & \multicolumn{1}{p{4.19em}}{Corrupted} & \multicolumn{1}{p{4.19em}}{~~DDA} & \multicolumn{1}{p{4.19em}|}{~~Ours} & \multicolumn{1}{p{4.19em}}{Corrupted} & \multicolumn{1}{p{4.19em}}{~~DDA} & \multicolumn{1}{p{4.19em}|}{~~Ours} & \multicolumn{1}{p{4.19em}}{Corrupted} & \multicolumn{1}{p{4.19em}}{~~DDA} & \multicolumn{1}{p{4.19em}|}{~~Ours} \\
    \midrule
    \multicolumn{1}{p{8.44em}|}{original} & 0     & \multicolumn{3}{c|}{66.54 } & \multicolumn{3}{c|}{71.20 } & \multicolumn{3}{c|}{73.29 } \\
    \midrule
    \multicolumn{1}{l|}{\multirow{5}[2]{*}{defocus blur}} & 1     & 66.52  & 66.50  & 66.25  & 69.42  & 69.90  & 70.07  & 72.03  & 72.54  & 72.93  \\
          & 2     & 65.75  & 66.15  & 66.25  & 67.71  & 68.99  & 69.38  & 71.85  & 71.91  & 72.14  \\
          & 3     & 64.59  & 64.79  & 64.92  & 65.56  & 67.02  & 67.63  & 70.05  & 70.63  & 71.04  \\
          & 4     & 63.78  & 64.06  & 64.27  & 64.47  & 65.42  & 66.25  & 68.53  & 69.48  & 69.80  \\
          & 5     & 63.39  & 63.74  & 63.98  & 62.97  & 63.58  & 63.94  & 66.70  & 67.25  & 68.02  \\
    \midrule
    \multicolumn{1}{l|}{\multirow{5}[2]{*}{glass blur}} & 1     & 66.46  & 66.29  & 66.15  & 70.78  & 71.02  & 70.31  & 72.44  & 72.99  & 72.89  \\
          & 2     & 66.05  & 66.11  & 66.09  & 69.52  & 70.09  & 69.97  & 72.18  & 72.76  & 72.72  \\
          & 3     & 65.40  & 65.61  & 65.69  & 68.75  & 69.22  & 69.58  & 71.87  & 71.95  & 72.14  \\
          & 4     & 64.81  & 65.06  & 65.08  & 67.65  & 68.18  & 68.67  & 71.06  & 71.36  & 71.51  \\
          & 5     & 64.02  & 64.00  & 64.31  & 65.69  & 66.42  & 66.68  & 69.15  & 69.90  & 70.35  \\
    \midrule
    \multicolumn{1}{l|}{\multirow{5}[2]{*}{motion blur}} & 1     & 66.50  & 66.42  & 66.25  & 70.59  & 70.41  & 70.05  & 72.66  & 73.23  & 73.13  \\
          & 2     & 66.13  & 66.32  & 66.42  & 69.07  & 69.48  & 69.78  & 72.05  & 72.42  & 72.50  \\
          & 3     & 65.26  & 65.44  & 65.42  & 67.61  & 68.40  & 68.57  & 71.06  & 71.69  & 71.51  \\
          & 4     & 64.37  & 64.71  & 65.08  & 66.29  & 66.94  & 67.45  & 69.30  & 69.82  & 70.07  \\
          & 5     & 63.90  & 64.17  & 64.23  & 65.56  & 66.19  & 66.07  & 67.90  & 68.46  & 68.97  \\
    \midrule
    \multicolumn{1}{l|}{\multirow{5}[2]{*}{zoom blur}} & 1     & 66.31  & 66.54  & 66.60  & 70.70  & 70.57  & 69.99  & 73.03  & 72.95  & 73.25  \\
          & 2     & 65.93  & 66.23  & 66.19  & 69.97  & 70.31  & 69.64  & 72.50  & 72.58  & 72.87  \\
          & 3     & 65.59  & 65.67  & 65.63  & 69.01  & 69.24  & 68.53  & 71.97  & 71.95  & 72.05  \\
          & 4     & 64.88  & 65.34  & 65.36  & 68.34  & 68.97  & 68.42  & 71.41  & 72.01  & 72.12  \\
          & 5     & 64.23  & 64.59  & 64.67  & 67.82  & 68.14  & 67.17  & 70.33  & 70.68  & 70.27  \\
    \midrule
    \multicolumn{1}{l|}{\multirow{5}[2]{*}{gaussian blur}} & 1     & 66.48  & 66.34  & 66.40  & 70.90  & 71.20  & 71.24  & 72.95  & 73.11  & 73.21  \\
          & 2     & 65.95  & 66.32  & 66.32  & 68.34  & 69.42  & 69.80  & 72.01  & 72.30  & 72.34  \\
          & 3     & 64.55  & 64.73  & 65.24  & 66.03  & 67.51  & 67.88  & 70.25  & 70.94  & 71.55  \\
          & 4     & 63.90  & 64.35  & 64.49  & 64.47  & 65.95  & 66.56  & 69.13  & 69.46  & 69.86  \\
          & 5     & 62.91  & 63.33  & 63.13  & 61.77  & 62.60  & 62.62  & 65.16  & 66.05  & 66.60  \\
    \midrule
    \multicolumn{1}{l|}{\multirow{5}[2]{*}{elastic transform}} & 1     & 67.05  & 66.80  & 66.44  & 70.90  & 70.66  & 70.43  & 72.38  & 72.66  & 72.38  \\
          & 2     & 65.54  & 65.54  & 65.30  & 70.45  & 70.03  & 70.43  & 70.61  & 70.57  & 70.70  \\
          & 3     & 66.64  & 66.68  & 66.13  & 71.12  & 71.10  & 70.51  & 72.87  & 73.13  & 72.83  \\
          & 4     & 66.88  & 66.70  & 66.27  & 71.10  & 71.24  & 70.47  & 73.07  & 73.05  & 72.60  \\
          & 5     & 66.62  & 66.58  & 66.40  & 71.34  & 71.28  & 70.37  & 72.68  & 72.83  & 72.36  \\
    \midrule
    \multicolumn{1}{l|}{\multirow{5}[2]{*}{jpeg compression}} & 1     & 66.23  & 65.93  & 65.69  & 69.52  & 68.61  & 67.63  & 70.29  & 70.53  & 70.15  \\
          & 2     & 66.21  & 65.87  & 65.50  & 67.27  & 66.86  & 65.79  & 69.21  & 69.48  & 69.26  \\
          & 3     & 65.61  & 65.50  & 65.30  & 65.81  & 64.86  & 64.21  & 67.37  & 67.45  & 67.57  \\
          & 4     & 62.56  & 62.81  & 62.85  & 60.78  & 59.62  & 58.67  & 61.71  & 62.26  & 62.30  \\
          & 5     & 56.72  & 56.95  & 57.03  & 55.81  & 55.45  & 55.47  & 57.57  & 57.80  & 58.10  \\
    \midrule
    \multicolumn{1}{l|}{\multirow{5}[2]{*}{pixelate}} & 1     & 66.60  & 66.46  & 66.15  & 71.32  & 71.24  & 70.51  & 73.41  & 73.68  & 73.60  \\
          & 2     & 66.58  & 66.54  & 66.40  & 71.49  & 71.09  & 70.65  & 73.41  & 73.45  & 73.19  \\
          & 3     & 66.38  & 66.29  & 65.95  & 70.88  & 71.04  & 70.86  & 72.83  & 73.09  & 73.13  \\
          & 4     & 66.38  & 66.42  & 66.29  & 70.45  & 70.63  & 69.92  & 72.50  & 73.15  & 72.95  \\
          & 5     & 66.76  & 66.56  & 66.30  & 61.49  & 68.12  & 70.41  & 70.23  & 71.87  & 72.82  \\
    \midrule
    \multicolumn{1}{l|}{\multirow{5}[2]{*}{gaussian noise}} & 1     & 57.82  & 60.74  & 62.83  & 64.00  & 67.75  & 69.40  & 69.50  & 70.49  & 71.20  \\
          & 2     & 45.73  & 51.98  & 57.15  & 59.16  & 64.21  & 68.14  & 64.15  & 66.46  & 68.87  \\
          & 3     & 30.70  & 38.17  & 43.95  & 54.72  & 61.00  & 65.26  & 57.68  & 61.65  & 64.51  \\
          & 4     & 25.80  & 25.61  & 26.26  & 51.53  & 54.46  & 61.14  & 50.96  & 56.07  & 60.07  \\
          & 5     & 24.54  & 21.88  & 19.31  & 50.68  & 51.69  & 55.59  & 49.38  & 51.77  & 54.65  \\
    \midrule
    \multicolumn{1}{l|}{\multirow{5}[2]{*}{impulse noise}} & 1     & 55.16  & 59.14  & 62.06  & 64.67  & 67.41  & 68.71  & 68.34  & 69.84  & 70.45  \\
          & 2     & 40.52  & 49.12  & 54.90  & 60.41  & 64.39  & 66.36  & 62.97  & 65.65  & 67.67  \\
          & 3     & 31.86  & 39.67  & 46.22  & 57.23  & 61.91  & 64.96  & 58.81  & 62.28  & 64.63  \\
          & 4     & 23.08  & 24.03  & 25.96  & 52.30  & 56.60  & 60.60  & 50.70  & 55.59  & 59.32  \\
          & 5     & 21.38  & 20.12  & 18.84  & 50.72  & 52.42  & 55.93  & 48.85  & 51.33  & 55.18  \\
    \midrule
    \multicolumn{1}{l|}{\multirow{5}[2]{*}{shot noise}} & 1     & 52.54  & 57.45  & 61.18  & 61.77  & 65.95  & 69.17  & 66.62  & 68.97  & 70.63  \\
          & 2     & 38.19  & 45.63  & 51.49  & 57.76  & 61.98  & 66.46  & 61.31  & 64.57  & 67.07  \\
          & 3     & 29.37  & 32.51  & 37.03  & 54.78  & 58.61  & 63.41  & 55.49  & 59.14  & 62.68  \\
          & 4     & 27.26  & 24.76  & 21.90  & 52.20  & 55.38  & 58.87  & 49.89  & 53.86  & 56.86  \\
          & 5     & 28.03  & 22.27  & 18.86  & 51.17  & 53.28  & 54.94  & 49.32  & 52.00  & 53.46  \\
    \midrule
    \multicolumn{1}{l|}{\multirow{5}[2]{*}{speckle noise}} & 1     & 55.02  & 59.76  & 62.38  & 63.25  & 66.52  & 68.89  & 68.38  & 70.11  & 70.63  \\
          & 2     & 45.87  & 52.73  & 57.76  & 60.07  & 64.61  & 67.84  & 64.98  & 67.49  & 68.91  \\
          & 3     & 30.99  & 35.15  & 39.10  & 56.24  & 59.16  & 64.06  & 57.51  & 60.23  & 63.13  \\
          & 4     & 29.49  & 29.67  & 28.55  & 54.84  & 57.31  & 61.37  & 54.17  & 57.17  & 60.33  \\
          & 5     & 29.69  & 26.51  & 22.15  & 53.13  & 55.24  & 57.55  & 51.96  & 54.53  & 56.42  \\
    \midrule
    \multicolumn{1}{l|}{\multirow{5}[2]{*}{brightness}} & 1     & 62.16  & 61.89  & 61.65  & 70.80  & 70.45  & 69.80  & 72.22  & 72.12  & 72.07  \\
          & 2     & 55.42  & 55.61  & 55.63  & 69.09  & 68.57  & 67.88  & 70.29  & 70.33  & 69.92  \\
          & 3     & 51.06  & 51.15  & 51.21  & 67.15  & 67.07  & 66.54  & 68.63  & 68.69  & 68.55  \\
          & 4     & 48.37  & 48.65  & 48.92  & 65.18  & 65.08  & 64.73  & 66.48  & 66.82  & 66.74  \\
          & 5     & 47.11  & 47.25  & 47.46  & 63.29  & 63.25  & 62.93  & 65.20  & 65.14  & 65.26  \\
    \midrule
    \multicolumn{1}{l|}{\multirow{5}[2]{*}{spatter}} & 1     & 65.77  & 65.77  & 65.38  & 70.59  & 70.47  & 70.03  & 72.85  & 72.82  & 72.76  \\
          & 2     & 52.63  & 52.91  & 51.57  & 64.02  & 62.91  & 60.68  & 69.30  & 68.38  & 66.92  \\
          & 3     & 48.29  & 47.92  & 47.80  & 56.18  & 53.76  & 50.98  & 63.17  & 61.95  & 59.78  \\
          & 4     & 56.52  & 56.68  & 57.01  & 61.29  & 61.25  & 61.00  & 63.82  & 62.97  & 62.10  \\
          & 5     & 52.61  & 53.26  & 53.60  & 58.37  & 57.76  & 56.74  & 59.54  & 59.20  & 58.83  \\
    \bottomrule
    \end{tabular}%
\end{table}%

\end{document}